\documentclass[10pt,letterpaper,twocolumn]{article} %% two column, final layout

\usepackage{ol2}
\usepackage[draft]{hyperref}
\usepackage{amsmath}

\begin{document}

\twocolumn[ %% activate for two-column option

\title{$\mathcal{PT}$ phase control in circular multicore fibers}

%% For REVTeX it is possible to automate superscript and e-mail callouts with the superscriptaddress option; see REVTeX4 documentation.

\author{Stefano Longhi}

\address{Dipartimento di Fisica, Politecnico di Milano and Istituto di Fotonica e Nanotecnologie del Consiglio Nazionale delle Ricerche, Piazza L. da Vinci 32, I-20133 Milano, Italy (stefano.longhi@polimi.it)}

\begin{abstract}
We consider light dynamics in a circular  multicore fiber with balanced gain and loss core distribution, and show that transition from unbroken to broken $\mathcal{PT}$ phases can be conveniently controlled by geometric twist of the fiber. The twist introduces Peierls' phases in the coupling constants and thus acts as an artificial gauge field. As an application, we discuss twist-induced tuning of optical transmission in a six-core fiber with one lossy core. 
\end{abstract}

\ocis{070.7345, 130.0130, 060.2310}
 ] %% activate for two-column option

Parity-time ($\mathcal{PT}$) symmetry, originated in the context of quantum mechanics
and quantum field theory \cite{r1}, has inspired in the past recent years several new ideas in optics \cite{r2,r3,r4,r5,r6,r7,r8,r9,r10}, with 
potential applications to optical beam engineering, image
processing, asymmetric transmission, laser mode selection, mode conversion, etc. \cite{r10,r11,r12,r13,r14,r15,r16,r17,r18}. A  property of $\mathcal{PT}$-symmetric optical structures is
the existence of two distinguished phases, e.g., one with
real energy spectrum (unbroken $\mathcal{PT}$ symmetry) and the
other one with complex conjugate eigenvalues (broken $\mathcal{PT}$
symmetry) corresponding to the onset of optical instabilities \cite{r19}. In periodically-modulated systems, broken and unbroken $\mathcal{PT}$ phases are determined by Floquet quasi-energies \cite{r20,r21,r22,r23,r24,r25}. Phase transition is generally observed when the level of balanced gain and loss in the system is increased  above a threshold value, however exception to such a rule can be found \cite{r26}.  Among various optical structures, coupled  optical waveguides or microrings in integrated photonic platforms provide one of the most appealing and experimentally accessible setting to implement the idea of $\mathcal{PT}$-symmetry  in optics. 
An important issue in such structures is the ability to control and tune the $\mathcal{PT}$ phase transition. A possibility is obviously to vary the level of gain and loss in the system, however this is a rather challenging task since gain and loss should remain balanced and thus must be tuned simultaneously. Geometry can provide a useful and simple means to this aim. For example, in Ref.\cite{r27} it was shown that the introduction of a strain in a $\mathcal{PT}$-symmetric honeycomb lattice can restore the $\mathcal{PT}$-symmetric phase.\\
In this Letter we consider light dynamics in a circular multicore fiber with balanced gain and loss distribution, and show that transition from unbroken to broken $\mathcal{PT}$ symmetric phases can be conveniently controlled by a geometric twist of the fiber. The twist introduces additional Peierls' phases in the coupling constants among the fiber cores \cite{r28,r29} and thus acts as an artificial gauge field which modifies wave transport in the structure \cite{r30,r31}. We consider specifically phase transition control in a $\mathcal{PT}$ multicore fiber comprising three (trimer) and six cores.  As an application, we discuss twist-induced tuning of optical transmission in a six-core fiber with one lossy core.\par
\begin{figure}[htb]
\centerline{\includegraphics[width=8cm]{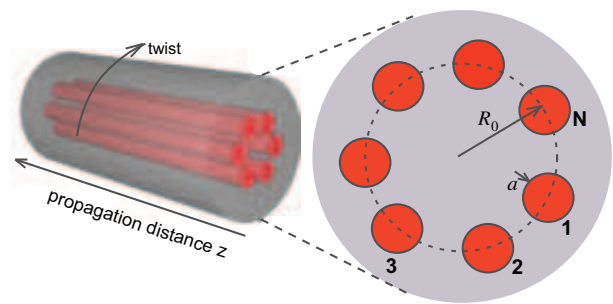}} \caption{ \small
(Color online) Schematic of a multicore fiber, comprising $N$ equal cores of radius $a$ uniformly distributed along a ring of radius $R_0$. The fiber is twisted along the propagation distance $z$ with a twist period $\Lambda$.}
\end{figure}

Let us consider a multicolore circular fiber comprising $N$ equal cores of radius $a$ equally spaced along a ring of radius $R_0$ (Fig.1), and let us assume that optical gain and loss is provided in the various cores of the fiber with a (typically) balanced distribution \cite{r32,r33}. In the absence of the twist and within a tight-binding model with nearest-neighbor couplings, light dynamics in the optical structure along the propagation axis $z$ is described by the following coupled-mode equations \cite{r32,r33}
\begin{equation}
i \frac{dc_n}{dz}=\kappa (c_{n+1}+c_{n-1})+i \gamma_n c_n
\end{equation}
($n=1,2,...,N$), where $c_n$ is the amplitude of the electric field in the $n-th$  waveguide core, $\kappa$ is the coupling constant of adjacent cores, $\gamma_n$ is the optical gain ($\gamma_n>0$) or loss ($\gamma_n<0$) rate in the $n-th$  core, and the cyclic periodic boundary conditions 
\begin{equation}
c_{n+N}=c_n
\end{equation}
have been assumed in writing Eq.(1). We note that for an even number $N$ of cores with alternating balanced gain and loss terms, i.e. $\gamma_n=-(-1)^n \gamma$, $\mathcal{PT}$ symmetry breaking threshold has been calculated in Ref.\cite{r32}; in particular, when $N/2$ is even the optical structure is always in the broken $\mathcal{PT}$ phase for any arbitrarily small value of $\gamma$.\\ 
When the fiber is twisted along the propagation axis $z$ with a uniform twist rate $\epsilon= 2 \pi / \Lambda$ and spatial period $\Lambda$, coupled-mode equations (1) are modified by the introduction of an additional Peierls' phase $\phi$ in the coupling constant \cite{r29}, namely one has
\begin{equation}
i \frac{dc_n}{dz}=\kappa{'}  \exp(-i \phi) c_{n+1} + \kappa{'} \exp(i \phi) c_{n-1}+i \gamma_n c_n
\end{equation}
where 
\begin{equation}
\phi=\frac{4 \pi^2 \epsilon n_s r_0^2}{N \lambda}
\end{equation}
In Eq.(4), $\lambda$ is the wavelength of the propagating field, $n_s$ is the substrate (cladding) refractive index, and $r_0 \simeq R_0$ is an effective radius, which is defined in Ref.\cite{r29} and typically turns out to be slightly smaller than $R_0$. The coupling constant $\kappa'=\kappa'(\epsilon)$ in the twisted fiber is usually slightly smaller than $\kappa$ and decreases as $\epsilon$ is increased because of the effect of the centrifugal force that pushes the light mode toward the outer region of the core \cite{r29}. However, at first approximation and for small twist rates one can assume $\kappa' \simeq \kappa$ in Eq.(3). The main effect of the Peierls' phase is to complexify the coupling constant, which in a different coupling scheme (so-called active coupling) was shown to enhance the stability domain (unbroken $\mathcal{PT}$ phase) \cite{referee}.\\
 \begin{figure}[htb]
\centerline{\includegraphics[width=8.4cm]{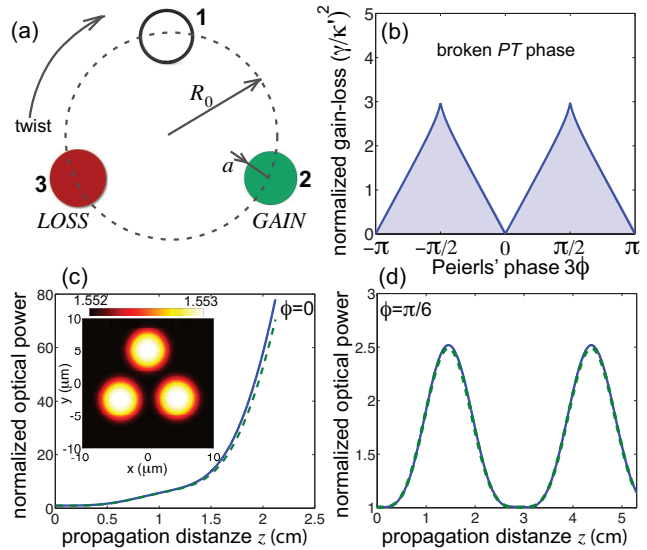}} \caption{ \small
(Color online) (a) Schematic of a three-core fiber with balanced gain and loss region in guides 2 and 3 ($\mathcal{PT}$-symmetric trimer). (b) Domains of unbroken and broken $\mathcal{PT}$ phase in the $(\phi, \gamma / \kappa')$ plane. The shaded regions correspond to unbroken $\mathcal{PT}$ phase. (c,d) Numerically-computed evolution of beam power in a three-core fiber (solid curves: full beam-propagation analysis, dashed curves: coupled-mode theory) for (c) untwisted, and (d) twisted fiber corresponding to a Peierls' phase $\phi= \pi/6$. Parameter values are given in the text. The fiber is initially excited in the fundamental mode of core 1. The inset in (c) shows the refractive index profile (real part) of the multicore fiber used in the full numerical beam propagation analysis.}
\end{figure}
To study the impact of fiber twist on the $\mathcal{PT}$ symmetry breaking transition,  let us first consider the simplest case of a $\mathcal{PT}$-symmetric trimer \cite{r34}, i.e. $N=3$, with $\gamma_1=0$, $\gamma_2=\gamma$ and $\gamma_3=-\gamma$ [Fig.2(a)]. In this case, the eigenvalues $E$ of Eq.(3), $c_n(z) \sim \exp(-iEz)$, are the roots of the cubic equation
\begin{equation}
E^3+E(\gamma^2-3 \kappa'^2)-2 \kappa'^3 \cos( 3 \phi)=0
\end{equation} 
which are given by (according to Cardano's formula)
\begin{subequations}
\begin{eqnarray}
E_1 & = & \kappa' (S_++S_-)\\
E_2 & = & -\frac{\kappa'}{2} (S_{+}+S_{-})+i \frac{\sqrt{3} \kappa'}{2}(S_+-S_-) \\
E_3 & = & -\frac{\kappa'}{2} (S_++S_-)-i \frac{\sqrt{3} \kappa'}{2}(S_+-S_-) 
\end{eqnarray}
\end{subequations}
where we base set 
 \begin{equation}
 S_{\pm}= \left\{ \cos( 3 \phi) \pm \sqrt{  \left[ \frac{1}{3} \left( \frac{\gamma}{\kappa'} \right)^2 -1 \right]^3  + \cos^2(3 \phi)} \right\}^{1/3}
 \end{equation}
Broken $\mathcal{PT}$ phase correspond to $S_{\pm}$ real, with $E_{2,3}$ complex conjugate energies, whereas unbroken $\mathcal{PT}$ phase corresponds to $S_{\pm}$ complex conjugates and all eigenvalues real according to Eq.(6). Unbroken $\mathcal{PT}$ phase is thus obtained when the condition
\begin{equation}
\left( \frac{\gamma}{\kappa'}\right)^2 < 3 \sqrt{1- \cos^{2/3} (3 \phi) }
\end{equation}
is met. Figure 2(b) shows the regions of broken/unbroken $\mathcal{PT}$ phases in the $(\phi, \gamma/ \kappa')$ plane. Note that in the untwisted fiber ($\phi=0$) the trimer is always in the broken $\mathcal{PT}$ phase, whereas fiber twist introduces a non vanishing symmetry breaking threshold, which is highest and equal to $\gamma/\kappa'= \sqrt{3}$ at $ \phi=\pi/6$, i.e. for a twits rate 
\begin{equation}
\epsilon=\frac{\lambda}{8 \pi n_s r_0^2}
\end{equation}
Therefore fiber twist can restore $\mathcal{PT}$ symmetry. This is shown as an example in Figs.2(c) and (d). The figures show the numerically-computed evolution of the optical power, normalized to its input value, in a three-core step-index optical fiber probed at $\lambda=980$ nm with radius $R_0=5 \; \mu$m, cladding refractive index $n_s=1.552$, core radius $a=3 \; \mu$m, and core refractive index change $\Delta n= \Delta n_R+i \Delta n_I$, with $\Delta n_R=0.01$, $\Delta n_I=0$ in core 1  and $\Delta n_I=\pm 0.004 \times \Delta n_R$ in cores 2 and 3. The fiber is initially excited in the fundamental mode of core 1, and power evolution in shown for the untwisted fiber [Fig.2(c)] and for a twisted fiber [Fig.2(d)] with twist period $\Lambda =2.56 \; {\rm mm}$, corresponding to a Peierls' phase $\phi \simeq \pi/6$ (taking $r_0 \simeq 0. 65 R_0$). The solid curves refer to full numerical analysis, obtained by solving the paraxial Schr\"{o}dinger-like optical wave equation in the twisted fiber using a standard pseudo spectral split-step method (see, for instance, \cite{r28,r29});  dashed curves refer to the predictions of coupled-mode theory [Eqs.(3)] with $\kappa=1.95 \; \rm{cm}^{-1}$, $\gamma=2.04 \; \rm{cm}^{-1}$, $\phi=0$ in Fig.2(c), and $\kappa'=1.71 \; \rm{cm}^{-1}$, $\gamma=2.04 \; \rm{cm}^{-1}$ and $\phi= \pi/6$ in Fig.2(d). Note that, while in the untwisted fiber a secular growth of optical power is observed, which is the signature of a broken $\mathcal{PT}$ phase, in the twisted fiber the $\mathcal{PT}$ symmetry is restored and the optical power is oscillatory. It should be noted that the Peierls' phase introduces asymmetric (chiral) dynamics for clockwise/counterclockwise fiber rotation, however symmetry breaking threshold does not depend on the rotation side.\\
As a second case, let us consider a multicore fiber with an even number $N=2M$ of cores with alternating gain and loss region, i.e. $\gamma_n=-(-1)^n \gamma$ in Eq.(3). Light dynamics in such a structure in the absence of the Peierls' phase ($\phi=0$) was investigated in Ref.\cite{r32}. In particular, when $M$ is even $\mathcal{PT}$ symmetry is always broken, whereas when $M$ is odd  $\mathcal{PT}$ symmetry is unbroken for a sufficiently small value of $\gamma/ \kappa'$. For $\phi \neq 0$, the $N=2M$ eigenvalues $E=E_l$ of Eq.(3) can be readily calculated by the Ansatz $c_n(z)=\exp(iqn-iEz)$, where the allowed values of the Bloch wave number $q$ are constrained by the periodicity condition $c_{n+N}=c_n$. This yields 
\begin{equation}
E_l= \pm \sqrt{4 \kappa'^2 \cos^2(q_l+\phi)-\gamma^2} 
\end{equation} 
 \begin{figure}[htb]
\centerline{\includegraphics[width=8.4cm]{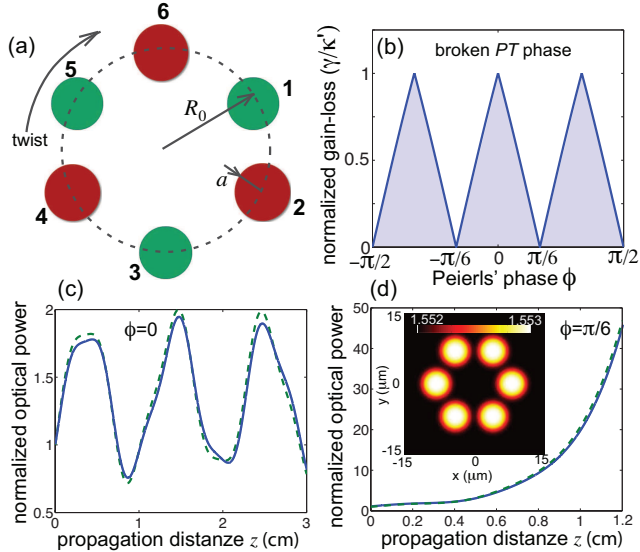}} \caption{ \small
(Color online) (a) Schematic of a six-core fiber with alternating gain/loss cores. (b) Domains of unbroken and broken $\mathcal{PT}$ phases in the $(\phi, \gamma / \kappa')$ plane. The shaded regions correspond to unbroken $\mathcal{PT}$ phase. (c,d) Numerically-computed evolution of beam power in the six-core fiber (solid curves: full beam-propagation analysis, dashed curves: coupled-mode theory) for (c) untwisted, and (d) twisted fiber corresponding to a Peierls' phase $\phi= \pi/6$. Parameter values are given in the text. The fiber is initially excited in the fundamental mode of core 1. The inset in (d) shows the refractive index profile (real part) of the multicore fiber used in the full numerical beam propagation analysis.}
\end{figure}
with $q_l=l \pi/M$ ($l=0,1,2,...,M-1$). As it can be seen, the Peierls' phase shifts the Bloch wave number ($q_l \rightarrow q_l + \phi$) in the dispersion relation (10) and generally changes the symmetry breaking transition point. Let us focus our attention to the six-core fiber ($N=6$) [Fig.3(a)], which shows some interesting dynamical behavior. The $\mathcal{PT}$ phase diagram of the six-core fiber in the $(\gamma/ \kappa', \phi)$ plane, as obtained from an inspection of Eq.(10), is shown in Fig.3(b). Note that in the untwisted fiber ($\phi=0$) $\mathcal{PT}$ symmetry breaking is observed when $\gamma / \kappa'$ is increased above 1, whereas for a twisted fiber with Peierls' phase $\phi= \pi/6$, corresponding to a twist rate
\begin{equation}
\epsilon=\frac{\lambda}{4 \pi n_s r_0^2}
\end{equation}
 \begin{figure}[htb]
\centerline{\includegraphics[width=8.4cm]{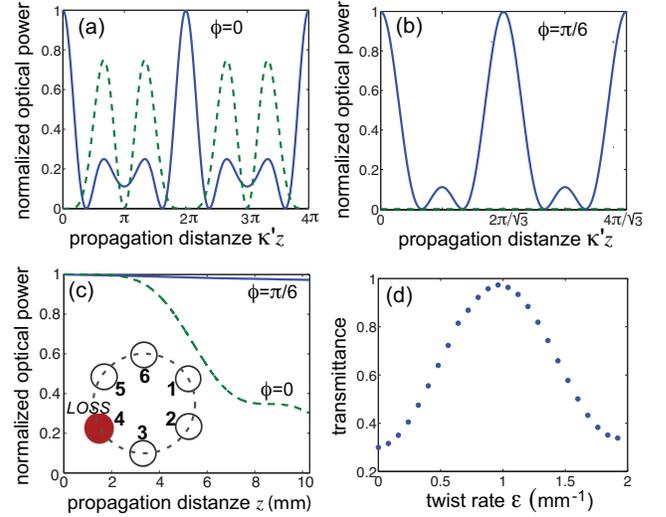}} \caption{ \small
(Color online) (a,b) Light dynamics in a six-core fiber without  loss/gain, as obtained from Eq.(3) with $\gamma_n=0$, for (a) $\phi=0$ and (b) $\phi=\pi/6$ with initial condition $c_l(0)= \delta_{l,1}$ (excitation of core 1). The solid (dashed) curve shows the evolution of normalized optical power trapped in core 1 (core 4). (c) Evolution of normalized optical power in a 10.3-mm-long six-core optical fiber with loss in core 4, as obtained by beam propagation analysis. The refractive index profile (real part) of the fiber is the same as in the inset of Fig.3(d). The fiber is excited at the input plane in the fundamental mode of core 1. The dashed curve refers to the untwisted fiber, whereas the solid curve to a twisted fiber with twist period $\Lambda=6.54$ mm, corresponding to a Peierls' phase $\phi \simeq \pi/6$. (d) Numerically-computed behavior of fiber transmittance versus the twist rate $\epsilon$ (fiber length 10.3 mm).}
\end{figure}
$ \mathcal{PT}$ symmetry is always broken. This means that, as opposed to the three-core fiber discussed above, in this case fiber twist can induce $\mathcal{PT}$ symmetry breaking. An example of twist-induced  $\mathcal{PT}$ symmetry breaking is shown in Figs.3(c) and (d). The figures show the numerically-computed evolution of the optical power, normalized to its input value, in the six-core   fiber probed at $\lambda=980$ nm with radius $R_0=8 \; \mu$m, cladding refractive index $n_s=1.552$, core radius $a=3 \; \mu$m, and core refractive index change $\Delta n= \Delta n_R+i \Delta n_I$, with $\Delta n_R=0.01$ and $\Delta n_I=\pm 0.004 \times \Delta n_R$ at alternating cores. The fiber is initially excited in the fundamental mode of core 1, and power evolution is shown for the untwisted fiber [Fig.3(c)] and for a twisted fiber [Fig.3(d)] with twist period $\Lambda =6.54 \; {\rm mm}$, corresponding to a Peierls' phase $\phi \simeq \pi/6$ (taking $r_0 \simeq 0. 91 R_0$). The dotted curves in the figures refer to the predictions of coupled-mode theory [Eqs.(3)] with $\kappa=3.59 \; \rm{cm}^{-1}$, $\gamma=2.05 \; \rm{cm}^{-1}$, $\phi=0$ in Fig.3(c), and 
$\kappa' \simeq \kappa=3.59 \; \rm{cm}^{-1}$, $\gamma=2.05 \; \rm{cm}^{-1}$ and $\phi= \pi/6$ in Fig.3(d).\par
It should be noted that the interplay between Peierls' phase (geometric twist) and non-Hermitian dynamics (light amplification and attenuation) acts also in non-${\mathcal PT}$ invariant structures, for example in purely dissipative (lossy) fibers. As an example, we discuss twist-induced tuning of optical transmission in a six-core fiber with one lossy core. In this case, there is no gain regions and in Eq.(3) we assume $\gamma_n=0$ for $n \neq 1$ and $\gamma_1=-\gamma<0$ for $n=1$ (the lossy core).  To explain the idea of twist-induced tuning of optical transmission, let us first consider the dissipationless case $\gamma=0$. In this case and for the untwisted fiber ($\phi=0$), from Eq.(10) with $\phi=\gamma=0$ it follows that there are four distinct energies, namely $ \pm \kappa'$ and $ \pm 2 \kappa'$, resulting in a periodic dynamics with spatial period $z_T= 2 \pi / \kappa'$. Hence, if at $z=0$ the fiber is excited in core 1, after a propagation distance $z=z_T$ a full revival, with light returning in core 1, is observed [Fig.4(a)]. Note that the core 4, symmetrically placed with respect to core 1, is excited during the dynamics [dashed curve in Fig.4(a)]. When the fiber is twisted with a Peierls' phase $\phi= \pi/6$, from Eq.(10) one obtains the three distinct eigenvalues $0$ and $\pm \sqrt{3} \kappa'$, resulting again in a periodic dynamics with modified period $z_T=2 \pi / ( \sqrt{3} \kappa')$.  Interestingly, when the core 1 is excited at plane $z=0$, after the propagation distance $z_T$ full revival is obtained [Fig.4(b)] like in the untwisted case, however now the core 4, symmetrically placed with respect to core 1, is never excited during the dynamics [dashed curve in Fig.4(b)], i.e. it is a dark core. Therefore, if some loss is now introduced in core 4 and the fiber is initially excited in core 1 [see inset in Fig.4(c)], we expect that optical power is conserved along the propagation in the twisted fiber despite of the lossy core, whereas in the untwisted fiber power attenuation should occur. Such a behavior is demonstrated in Fig.4(c). The figure depicts the numerically-computed evolution of the optical power along a 10.3-mm-long six-core fiber with the same parameters as in Fig.3, except than the imaginary part of the refractive index $\Delta n_I$ is introduced in core 4 solely. In the untwisted case (dashed curve) only $\sim 30 \%$ of the coupled optical power is transmitted, whereas for a twisted fiber with twist period $\Lambda=6.54$ mm, corresponding to a Peierls' phase $\phi= \pi/6$, about $97.4 \%$ of the coupled optical power is transmitted (solid curve). By changing the twist rate of the fiber, a variable optical attenuator can be therefore realized. This is shown in Fig.4(d), which depicts the optical transmittance of the fiber as a function of the twist rate $\epsilon= 2 \pi / \Lambda$. \par
In conclusion, we have theoretically investigated light dynamics in $\mathcal{PT}$-symmetric multicolor fibers and have shown the possibility to control the broken/unbroken $\mathcal{PT}$ phase transition by a uniform fiber twist. The geometric twist introduces Peierls' phases in the coupling constants of adjacent cores and thus acts as an artificial gauge field that modifies the tunneling dynamics  among the guiding cores. Geometric twist can be applied, for example, to the control of symmetry breaking transitions without resorting to gain/loss control, or for the realization of compact variable fiber attenuators. Our results disclose the major role  of artificial gauge fields on $\mathcal{PT}$ symmetry breaking transition and may stimulate further investigations, for example by including nonlinearities in the $\mathcal{PT}$-symmetric structures \cite{r32,r33,r34}. The interplay between $\mathcal{PT}$ symmetry and artificial gauge fields, which has been here studied for a multicore fiber model, could be also extended to other experimentally-accessible photonic structures, such as coupled resonators with synthetic magnetic fields in one and two-dimensional settings \cite{r35}.

\newpage

%%%%%%%%%%%%%%%%%%%%%%%%%%%%%%%
% References with full titles %
%%%%%%%%%%%%%%%%%%%%%%%%%%%%%%%

%\footnotesize
 {\bf References with full titles}\\
\\
1. C.M. Bender, {\it Making sense of non-Hermitian Hamiltonians}, Rep. Prog. Phys. {\bf 70}, 947 (2007).\\
2. A Ruschhaupt, F Delgado, and J.G. Muga, {\it Physical realization of $\mathcal{PT}$-symmetric potential scattering in a planar slab waveguide}, J. Phys. A {\bf 38}, L171 (2005).\\
3. R. El-Ganainy, K.G. Makris, D.N. Christodoulides, and Z.H. Musslimani, {\it Theory of coupled optical $\mathcal{PT}$-symmetric structures}, Opt. Lett. {\bf 32}, 2632  (2007).\\
4. K. G. Makris, R. El-Ganainy, D. N. Christodoulides, and Z. H. Musslimani, {\it Beam Dynamics in $\mathcal{PT}$-Symmetric Optical Lattices}, Phys. Rev. Lett. {\bf 100}, 103904 (2008).\\
5. S. Longhi, {\it Bloch Oscillations in Complex Crystals with $\mathcal{PT}$ Symmetry}, Phys. Rev. Lett. {\bf 103}, 123601 (2009).\\
6. C. E. R\"{u}ter, K.G. Makris, R. El-Ganainy, D.N. Christodoulides, M. Segev, and D. Kip, {\it Observation of parity-time symmetry in optics}, Nature Phys. {\bf 6}, 192 (2010).\\
7. S. Longhi, {\it $\mathcal{PT}$ symmetric laser-absorber}, Phys. Rev. A {\bf 82}, 031801 (2010).\\
8. Y.D. Chong, L. Ge, and A.D. Stone, {\it $\mathcal{PT}$-symmetry breaking and laser-absorber modes in optical scattering systems}, Phys. Rev. Lett. {\bf 106}, 093902 (2011).\\
9. A. Regensburger, C. Bersch, M.-A. Miri, G. Onishchukov, D.N. Christodoulides, and U. Peschel, {\it Parity-time synthetic photonic lattices}, Nature {\bf 488}, 167 (2012).\\
10. L. Feng, Y.-L. Xu, W.S. Fegadolli, M.-H. Lu, J.E.B. Oliveira, V.R. Almeida, Y.-F. Chen, and A. Scherer, {\it Experimental demonstration of a unidirectional reflectionless parity-time metamaterial at optical frequencies}, Nature Mat. {\bf 12}, 108 (2013).\\
11.F. Nazari, N. Bender, H. Ramezani, M. K.Moravvej-Farshi, D. N. Christodoulides, and T. Kottos, {\it Optical isolation via $\mathcal{PT}$-symmetric nonlinear Fano resonances}, Opt. Express {\bf 22}, 9575 (2014).\\
12. B. Peng, S. K. Ozdemir, F. Lei, F. Monifi, M. Gianfreda, G. L. Long, S. Fan, F. Nori, C. M. Bender, and L. Yang, {\it Parity-time-symmetric whispering-gallery microcavities}, Nature Phys. {\bf 10}, 394 (2014).\\
13. S. Longhi and L. Feng, {\it $\mathcal{PT}$-symmetric microring laser absorber}, Opt. Lett. {\bf 39}, 5026 (2014).\\
14. H. Hodaei, M.A. Miri, M. Heinrich, D.N. Christodoulides, and M. Khajavikhan, {\it Parity-time-symmetric microring lasers}, Science {\bf 346}, 975 (2014).\\
15. L. Feng, Z. J. Wong, R. M. Ma, Y. Wang, and X. Zhang, {\it Single-mode laser by parity-time symmetry breaking}, Science {\bf 346}, 972 (2014).\\
16. V.A. Vysloukh and Y.V. Kartashov, {\it Resonant mode conversion in the waveguides with unbroken and broken $\mathcal{PT}$ symmetry}, Opt. Lett. {\bf 39}, 5933 (2014).\\  
17. H. Hodaei, M. A. Miri, A. U. Hassan, W. E. Hayenga, M. Heinrich, D. N. Christodoulides, and M. Khajavikhan, {\it Parity-time-symmetric coupled microring lasers operating around an exceptional point}, Opt. Lett. {\bf 40}, 4955 (2015).\\
18. Y.V. Kartashov, V.A. Vysloukh, V.V. Konotop, and L. Torner, {\it Diffraction control in $\mathcal{PT}$-symmetric photonic lattices: From beam rectification to dynamic localization}, Phys. Rev. A {\bf 93}, 013841 (2016).\\
19. S. Longhi, {\it Optical Realization of Relativistic Non-Hermitian Quantum Mechanics}, Phys. Rev. Lett. {\bf 105}, 013903 (2010).\\
20. G. Della Valle and S. Longhi, {\it Spectral and transport properties of time-periodic $\mathcal{PT}$-symmetric tight-binding lattices}, Phys. Rev. A {\bf 87}, 022119 (2013).\\
21. X. Luo, J. Huang, H. Zhong, X. Qin, Q. Xie, Yu. S. Kivshar, and C. Lee, {\it Pseudo-Parity-Time Symmetry in Optical Systems}, Phys. Rev. Lett. {\bf 110}, 243902 (2013).\\
22. C. Yuce, {\it Pseudo PT symmetric lattice}, Eur. Phys. J. D {\bf 69}, 11 (2015).\\
23. J. Gong and Q.-H. Wang, {\it Stabilizing non-Hermitian systems by periodic driving}, Phys. Rev. A {\bf 91}, 042135 (2015).\\
24. S. Nixon and J. Yang, {\it Light propagation in periodically modulated complex waveguides}, Phys. Rev. A {\bf 91}, 033807 (2015).\\ 
25. G. Harari, Y. Plotnik, M. Bandres, Y. Lumer, M. C. Rechtsman, and M. Segev, {\it Topological insulators in $\mathcal{PT}$-symmetric lattices}, in CLEO: 2015, OSA Technical Digest (online) (Optical Society of America, 2015), paper FTh3D.3.\\
26. C.H. Liang, D.D. Scott, and Y.N. Joglekar, {\it $\mathcal{PT}$ restoration via increased loss and gain in the $\mathcal{PT}$-symmetric Aubry-Andr\'{e} model},
Phys. Rev. A {\bf 89}, 030102 (2014).\\
27. A. Szameit, M.C. Rechtsman, O. Bahat-Treidel, and M. Segev, {\it $\mathcal{PT}$-symmetry in honeycomb photonic lattices}, Phys. Rev. A {\bf 84}, 021806(R) (2011).\\
28. S. Longhi, {\it Bloch dynamics of light waves in helical optical waveguide arrays}, Phys. Rev. B {\bf 76}, 195119 (2007).\\
29. S. Longhi, {\it Light transfer control and diffraction management in circular fibre waveguide arrays}, J. Phys. B {\bf 40}, 4477 (2007).\\
30. X. Q. Li, X. Z. Zhang, G. Zhang, and Z. Song, {\it Asymmetric transmission through a flux-controlled non-Hermitian scattering center}, Phys. Rev. A {\bf 91}, 032101 (2015).\\
31. S. Longhi, {\it Non-reciprocal transmission in photonic lattices based on unidirectional coherent perfect absorption}, Opt. Lett. {\bf 40}, 1278 (2015).\\
32. I. V. Barashenkov, L. Baker, and N. V. Alexeeva, {\it $\mathcal{PT}$-symmetry breaking in a necklace of coupled optical waveguides}, Phys. Rev. A {\bf 87}, 033819 (2013).\\
33. A.J. Martinez, M.I. Molina, S.K. Turitsyn, and Yu. S. Kivshar, {\it Nonlinear multicore waveguiding structures with balanced gain and loss}, Phys. Rev. A {\bf 91}, 023822 (2015).\\
34. N. V. Alexeeva, I. V. Barashenkov, K. Rayanov, and S. Flach, {\it Actively coupled optical waveguides}, Phys. Rev. A {\bf 89}, 013848  (2014).\\
35. K. Li and P. G. Kevrekidis, {\it $\mathcal{PT}$-symmetric oligomers: Analytical solutions, linear stability, and nonlinear dynamics}, Phys. Rev. E {\bf 83}, 066608 (2011).\\
36. S. Mittal, S. Ganeshan, J. Fan, A. Vaezi, and M. Hafezi, {\it  Measurement of topological invariants in a 2D photonic system}, Nature Photon. {\bf 10}, 180 (2016).


\begin{thebibliography}{99}


%%%%%%%%%%%%%%%%%%%%%%%%%%%%%%%
% References (short version)  %
%%%%%%%%%%%%%%%%%%%%%%%%%%%%%%%

\bibitem{r1}
C.M. Bender, Rep. Prog. Phys. {\bf 70}, 947 (2007).
\bibitem{r2}
A Ruschhaupt, F Delgado, and J.G. Muga, J. Phys. A {\bf 38}, L171 (2005).
\bibitem{r3}
R. El-Ganainy, K.G. Makris, D.N. Christodoulides, and Z.H. Musslimani, Opt. Lett. {\bf 32}, 2632  (2007).
\bibitem{r4}
K. G. Makris, R. El-Ganainy, D. N. Christodoulides, and Z. H. Musslimani, Phys. Rev. Lett. {\bf 100}, 103904 (2008).
\bibitem{r5}
S. Longhi, Phys. Rev. Lett. {\bf 103}, 123601 (2009).
\bibitem{r6}
C. E. R\"{u}ter, K.G. Makris, R. El-Ganainy, D.N. Christodoulides, M. Segev, and D. Kip, Nature Phys. {\bf 6}, 192 (2010).
\bibitem{r7}
 S. Longhi, Phys. Rev. A {\bf 82}, 031801 (2010).
\bibitem{r8}
Y.D. Chong, L. Ge, and A.D. Stone, Phys. Rev. Lett. {\bf 106}, 093902 (2011).
\bibitem{r9}
A. Regensburger, C. Bersch, M.-A. Miri, G. Onishchukov, D.N. Christodoulides, and U. Peschel, Nature {\bf 488}, 167 (2012).
\bibitem{r10}
 L. Feng, Y.-L. Xu, W.S. Fegadolli, M.-H. Lu, J.E.B. Oliveira, V.R. Almeida, Y.-F. Chen, and A. Scherer, Nature Mat. {\bf 12}, 108 (2013).
\bibitem{r11}
F. Nazari, N. Bender, H. Ramezani, M. K.Moravvej-Farshi, D. N. Christodoulides, and T. Kottos, Opt. Express {\bf 22}, 9575 (2014).
\bibitem{r12}
B. Peng, S. K. Ozdemir, F. Lei, F. Monifi, M. Gianfreda, G. L. Long, S. Fan, F. Nori, C. M. Bender, and L. Yang, Nature Phys. {\bf 10}, 394 (2014).
\bibitem{r13}
S. Longhi and L. Feng,  Opt. Lett. {\bf 39}, 5026 (2014).
\bibitem{r14}
H. Hodaei, M.A. Miri, M. Heinrich, D.N. Christodoulides, and M. Khajavikhan, Science {\bf 346}, 975 (2014).
\bibitem{r15}
L. Feng, Z. J. Wong, R. M. Ma, Y. Wang, and X. Zhang, Science {\bf 346}, 972 (2014).
\bibitem{r16}
V.A. Vysloukh and Y.V. Kartashov,  Opt. Lett. {\bf 39}, 5933 (2014).
\bibitem{r17}
H. Hodaei, M. A. Miri, A. U. Hassan, W. E. Hayenga, M. Heinrich, D. N. Christodoulides, and M. Khajavikhan, Opt. Lett. {\bf 40}, 4955 (2015).
\bibitem{r18}
Y.V. Kartashov, V.A. Vysloukh, V.V. Konotop, and L. Torner, Phys. Rev. A {\bf 93}, 013841 (2016).
\bibitem{r19}
S. Longhi, Phys. Rev. Lett. {\bf 105}, 013903 (2010).
\bibitem{r20}
G. Della Valle and S. Longhi, Phys. Rev. A {\bf 87}, 022119 (2013).
\bibitem{r21}
X. Luo, J. Huang, H. Zhong, X. Qin, Q. Xie, Yu. S. Kivshar, and C. Lee, Phys. Rev. Lett. {\bf 110}, 243902 (2013).
\bibitem{r22}
C. Yuce, Eur. Phys. J. D {\bf 69}, 11 (2015).
\bibitem{r23}
J. Gong and Q.-H. Wang, Phys. Rev. A {\bf 91}, 042135 (2015).
\bibitem{r24}
S. Nixon and J. Yang, Phys. Rev. A {\bf 91}, 033807 (2015).
\bibitem{r25}
G. Harari, Y. Plotnik, M. Bandres, Y. Lumer, M. C. Rechtsman, and M. Segev,  CLEO: 2015, OSA Technical Digest (online) (Optical Society of America, 2015), paper FTh3D.3.
\bibitem{r26}
C.H. Liang, D.D. Scott, and Y.N. Joglekar, Phys. Rev. A {\bf 89}, 030102 (2014).
\bibitem{r27}
A. Szameit, M.C. Rechtsman, O. Bahat-Treidel, and M. Segev, Phys. Rev. A {\bf 84}, 021806(R) (2011).
\bibitem{r28}
S. Longhi, Phys. Rev. B {\bf 76}, 195119 (2007).
\bibitem{r29}
S. Longhi, J. Phys. B {\bf 40}, 4477 (2007).
\bibitem{r30}
X. Q. Li, X. Z. Zhang, G. Zhang, and Z. Song, Phys. Rev. A {\bf 91}, 032101 (2015).
\bibitem{r31}
S. Longhi, Opt. Lett. {\bf 40}, 1278 (2015).
\bibitem{r32}
I. V. Barashenkov, L. Baker, and N. V. Alexeeva, Phys. Rev. A {\bf 87}, 033819 (2013).
\bibitem{r33}
 A.J. Martinez, M.I. Molina, S.K. Turitsyn, and Yu. S. Kivshar, Phys. Rev. A {\bf 91}, 023822 (2015).
 \bibitem{referee}
 N. V. Alexeeva, I. V. Barashenkov, K. Rayanov, and S. Flach, Phys. Rev. A {\bf 89}, 013848  (2014).
\bibitem{r34}
K. Li and P. G. Kevrekidis, Phys. Rev. E {\bf 83}, 066608 (2011).
\bibitem{r35}
 S. Mittal, S. Ganeshan, J. Fan, A. Vaezi, and M. Hafezi, Nature Photon. {\bf 10}, 180 (2016).


\end{thebibliography}
\end{document}